\documentstyle[psfig]{l-aa}
\def\ltsima{$\; \buildrel < \over \sim \;$}
\def\lsim{\lower.5ex\hbox{\ltsima}}
\def\gtsima{$\; \buildrel > \over \sim \;$}
\def\gsim{\lower.5ex\hbox{\gtsima}}

\newcommand{\be}{\begin{equation}}
\newcommand{\en}{\end{equation}}
\def\deg {^\circ}
\def\ergs {~erg$\,$s$^{-1}$}

\def\cmdue {~cm$^{-2}$}

\def\aa #1 #2 {A\&A {#1} #2}
\def\aas #1 #2 {A\&AS {#1} #2}
\def\araa #1 #2 {ARA\&A {#1} #2}
\def\mon #1 #2 {MNRAS {#1} #2}
\def\apj #1 #2 {ApJ {#1} #2}
\def\apjs #1 #2 {ApJS {#1} #2}
\def\apjl #1 #2 {ApJ {#1} #2}
\def\aj #1 #2 {AJ {#1} #2}
\def\nat #1 #2 {Nature {#1} #2}
\def\pasj #1 #2 {PASJ {#1} #2}
\def\pasp #1 #2 {PASP {#1} #2}

\title{ X-ray variability in the quiescent state of Cen X-4}
\author{S.~Campana\inst{1,2}, S.~Mereghetti\inst{3}, L.~Stella\inst{4,2} \&
M.~Colpi\inst{5}}
                                 
\begin{document}

\institute{
{Osservatorio Astronomico di Brera, Via Bianchi 46, I-22055
Merate (Lc), Italy;}
\and
{Affiliated to I.C.R.A.}
\and
{Istituto di Fisica Cosmica del C.N.R., Via Bassini 15, I-20133 Milano,
Italy;}
\and
{Osservatorio Astronomico di Roma, Via dell'Osservatorio 2,
I-00040 Monteporzio Catone (Roma), Italy;} 
\and
{Dipartimento di Fisica, Universit\`a degli Studi di Milano,
Via Celoria 16, I-20133 Milano, Italy;} 
}
\offprints{Sergio Campana -- e-mail: campana@merate.mi.astro.it}

\date{Received: 5 December 1996; Accepted}

\maketitle
\label{sampout}

\begin{abstract}
We report on a ROSAT-HRI observation of the soft X--ray transient 
Cen X-4 during quiescence. 
We discover a variation in the flux by a factor of 3 in less than four days. 
This relatively fast variation, the first observed from a quiescent 
soft X--ray transient, rules out some of the emission 
mechanisms that have been proposed for the quiescent flux. 
Accretion either onto the neutron star surface or onto the magnetospheric
boundary is clearly favored.
\end{abstract}

\section{Introduction}
 
The quiescent emission of Soft X--ray Transients (SXRTs) 
has been investigated for only 5--6 sources (e.g., Verbunt et al. 1994; 
Asai et al. 1996); this is characterized in all cases by very soft spectra 
(black body temperatures 
of a few hundreds of eV) and X--ray luminosities of $\sim 10^{32-33}$\ergs. 

Cen X-4 is one of the nearest ($d\sim 1.2$ kpc) and best studied SXRT.
Two outbursts have been observed from Cen~X-4 in 1969 and 1979. 
The most detailed observation of Cen X-4 during quiescence is the one
recently obtained with ASCA by Asai et al. (1996).
These authors reported a quiescent luminosity of $2.4\times 10^{32}$\ergs\ 
(0.5--10 keV).
The X--ray spectrum measured is well fit by 
a black body component ($T_{\rm bb}=0.2$ keV; $R_{\rm bb}=1.8\times 10^6$ cm) 
plus a power-law with a photon index of 2--3. The contribution of these 
two spectral components in the ASCA energy band is comparable. 

Here we report on a ROSAT HRI observation of Cen X-4 during the quiescent period.

\section{ROSAT observation}

The ROSAT High Resolution Imager (HRI; David et al.
1992) observed Cen X-4 between 1995 August 16 and 26, for a net exposure time 
of 17469 s.
The brightest of the $\sim10$ X--ray sources detected in the field is at
coordinates 14h 58m 21s.8, $-31^{\deg} 40' 3''.6$ (J2000), fully compatible 
($<3''$) with the position of Cen X-4 reported by van Paradijs (1993).

The average HRI count rate of $(1.3\pm0.1)\times10^{-2}$ counts s$^{-1}$ 
corresponds to  a signal to noise ratio of 12.
By adopting the black body plus power law model derived from the ASCA fit 
(Asai et al. 1996) and a column density $N_H=6.6\times10^{20}$\cmdue\ 
as estimated by van Paradijs et al. (1987) based on the optical reddening,
we predict an HRI count rate of $\sim 10^{-2}$ counts s$^{-1}$.
This is in agreement with our measurement and indicates that the average 
source luminosity is at a level similar to that measured with ASCA $\sim 18$ 
months earlier.
For these spectral parameters and a distance of 1.2 kpc,
this count rate corresponds  to a 0.1--2.4 keV X--ray luminosity
of $\sim 7 \times 10^{31}$\ergs.

To study the source variability, we extracted 393 counts contained in 
a circle of $20''$ radius centered on the source  position. 
A periodicity search was carried out spanning periods from 10 ms to 100 s, 
with negative results.  
The corresponding upper limits are poor (pulsed fraction $\gsim 90\%$,
assuming a sinusoidal modulation) due 
to the high number of selected frequencies and the limited number of counts.
\begin{table}
\caption{ROSAT HRI Observation log.}
\label{tab1}
\begin{tabular}{ccrc}
Period begin & Period end & Exposure & Count rate \\
(TJD)        & (TJD)      & (s)\ \   & (counts s$^{-1}$)\\
\hline
9945.631 & 9945.648 & 1502 & $(6.8\pm0.7)\times10^{-2}$\\
9950.065 & 9950.480 & 4320& $(2.2\pm0.2)\times10^{-2}$\\
9953.589 & 9955.189 & 11647& $(1.6\pm0.1)\times10^{-2}$\\
\end{tabular}
\end{table}
\begin{figure}[htb]
\centerline{\psfig{figure=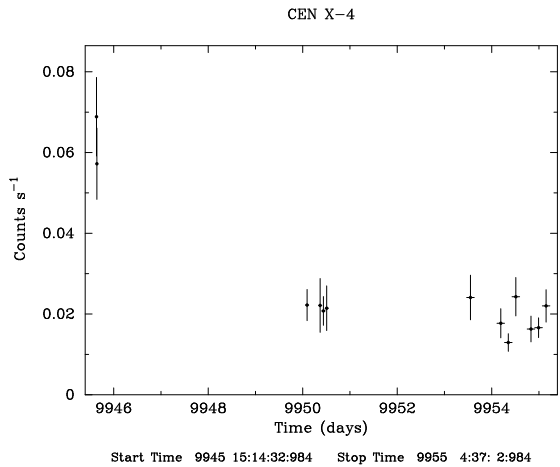,width=6cm}}
\caption{X--ray light curve of Cen X-4. Different bin times were applied 
for the three observing periods.}
\label{Fig1}
\end{figure}
If the search is restricted around the 32 Hz periodicity reported
by Mitsuda et al. (1996) based on the ASCA GIS
light curves, then an upper limit of 80\% on the pulsed fraction is
derived averaging together 6 power spectra of different time intervals, 
spanning 10 indipendent frequencies between 31.2498 and 31.2502 ms.

Significant aperiodic variability is clearly detected
on a timescale of days. Indeed, a
flux variation is apparent by comparing the Cen X-4
count rates obtained from the three separate time intervals
of which the ROSAT observation is composed (see Table 1 and Figure 1).
In the first interval a relatively high level of emission is observed, 
consistent with a flare-like event. This count rate variation
affects mainly the soft channels of the HRI.
In the other two intervals, separated 
by $\sim 4$ days each, the source count rate is a factor of 3--4 lower and 
its flux is consistent with a constant value. 

\section{Discussion}

In a reanalysis of the Einstein IPC (on 1980 July 28) and EXOSAT LE (on 1986 
February 21) 
observations of Cen X-4, we find that both are consistent with a single value 
of the X--ray luminosity for spectral parameters comparable to those of ASCA 
(Asai et al. 1996), hinting to a constant X--ray luminosity over 15 years.
Therefore the decrease in the X--ray luminosity found by van Paradijs et al.
(1987) assuming a 1 to 5 keV bremsstrahlung spectrum seems less likely.

Our ROSAT HRI observation of Cen X-4 provides also the first evidence 
for a flux variation on a timescale of a few days in the quiescent flux of 
a SXRT. This variation is likely intrinsic to the 
source (no occultations or dips have ever been observed) 
and therefore it can help constraining the emission mechanisms during 
quiescence (Stella et al. 1994).

The possibility that the late type companion star
accounts for the observed luminosity has been excluded by Verbunt (1996).
Neutron star cooling is attractive in view of the
spectral results consistent with thermal emission from a
body of size comparable with a neutron star, however
it is incompatible with the observed rapid 
decrease in the source intensity.
Also the shock emission model in which the X--ray luminosity is produced 
by the interaction of the relativistic pulsar wind with gaseous material 
from the companion star encounters problems. 
The mechanism is likely at work in PSR 1259--63, where the X--ray spectrum is 
characterized by a power-law with photon 
index of $\sim 2$ extending from 0.1 to 200 keV (e.g. Tavani \& Arons 1996).
In the case of Cen X-4 however, besides the power-law component, also a 
black body component is present. Moreover, the HRI count rate variation 
is mainly observed in the soft channels, where the black body
component dominates. 

We believe that the most likely explanation for the quiescent flux of Cen X-4
is that it derives from mass accretion. In this case variations
in the accretion flow can easily explain the observed changes. 
Either accretion onto the neutron star surface and onto the
magnetosphere can explain the observed properties, leaving open the 
possibility that Cen X-4 contains a fast spinning, 
weakly magnetized neutron star (Stella et al. 1994). 

\begin{acknowledgements}
This work has been partially supported through ASI grants.
\end{acknowledgements}

\end{document}